\documentclass[aps,pra,superscriptaddress,twocolumn,floatfix]{revtex4-1}
\usepackage[utf8]{inputenc}
\usepackage[T1]{fontenc}
\usepackage{amsmath}
\usepackage{amssymb}
\usepackage{mathtools}
\usepackage{graphicx}
\usepackage{overpic}
\usepackage[usenames,dvipsnames]{color}
\usepackage{bm}
\usepackage{array}
\usepackage{soul}
\sethlcolor{yellow}
\usepackage[colorlinks,urlcolor=blue,citecolor=blue,linkcolor=blue]{hyperref}
\usepackage{url}
\usepackage{subfigure}
\usepackage{slashed,bbm}
\usepackage{dsfont}
\usepackage{setspace}
\usepackage{soul}
\usepackage{braket}
\usepackage{sidecap}
\usepackage{multirow}
\usepackage{tensor}
\usepackage{ulem}
\usepackage{float}
\usepackage{booktabs}
\usepackage{color}

\begin{document}

\title{Random State Approach to Quantum Computation of Electronic-Structure Properties}

\author{Yiran Bai}
\affiliation{Institute of Advanced Technology, University of Science and Technology of China, Hefei 230000, China}
\affiliation{Yangtze Delta Industrial Innovation Center of Quantum Science and Technology, Suzhou 215000, China}
\author{Feng Xiong}
\email{xiongfeng@tgqs.net}
\affiliation{Yangtze Delta Industrial Innovation Center of Quantum Science and Technology, Suzhou 215000, China}
\author{Xueheng Kuang}
\email{xhkuang@whu.edu.cn}
\affiliation{Texas Materials Institute and Department of Mechanical Engineering, The University of Texas at Austin, Austin, Texas 78731, United States.}

\date{\today}
 
\begin{abstract}
Classical computation of electronic properties in large-scale materials remains challenging. Quantum computation has the potential to offer advantages in memory footprint and computational scaling. However, general and practical quantum algorithms for simulating large-scale materials are still lacking. We propose and implement random-state quantum algorithms to calculate electronic-structure properties of real materials. Using a random state circuit with only a few qubits, we employ real-time evolution with first-order Trotter decomposition and Hadamard test to obtain electronic density of states, and we develop a modified quantum phase estimation algorithm to calculate real-space local density of states via direct quantum measurements. Furthermore, we validate these algorithms by numerically computing the density of states and spatial distributions of electronic states in graphene, twisted bilayer graphene quasicrystals, and fractal lattices, covering system sizes from hundreds to thousands of atoms. Our results manifest that the random-state quantum algorithms provide a general and qubit-efficient route to simulating electronic properties of large-scale periodic and aperiodic materials on quantum computers.
\end{abstract}

\maketitle
\section{INTRODUCTION}
\label{sec:intro}
 Quantum computation was originally envisioned as a route to tackle quantum many-body problems, such as electron-electron interactions, more effectively than classical approaches by exploiting compact state representations and massive parallelism \cite{feynman1982simulating,lloyd1996universal}. In recent years, to simulate electronic ground-state properties in quantum computation, progress in relevant algorithmic has spanned near-term variational methods including the variational quantum eigensolver (VQE) and extensions\cite{peruzzo2014vqe,kandala2017hardware,grimsley2019adapt,tang2021qubitadapt}, quantum imaginary time evolution (QITE)\cite{motta2019qite,mcardle2019imaginary}, and quantum-Lanczos\cite{kirby2023qlanczos}, as well as fault-tolerant approaches based on quantum phase estimation (QPE)\cite{abrams1999qpe}, block-encoding\cite{rall2020quantum}, qubitization\cite{low2019qubitization}, and quantum signal processing \cite{low2017optimal,berry2015taylor,childs2012lcu}. For electronic excited states, subspace techniques\cite{nakanishi2019ssvqe}, such as the quantum equation of motion (qEOM)\cite{ollitrault2020qeom} and quantum subspace expansion (QSE)\cite{mcclean2017qse} , enabled the calculation of spectra and transition properties of small atoms and molecules on current devices. Algorithms targeting periodic solids and band structures are also emerging, including variational band-structure solvers, reduced-dimensional qEOM variants and full quantum eigensolver (FQE) \cite{sherbert2022band,fan2021band,zhang2024reducedqeom,babbush2018lowdepth,su2021faulttolerant,wei2020fqe,wang2024pfqe,wen2024fqess}. Nevertheless, practical approaches for computing electronic properties of large-scale periodic and nonperiodic materials on quantum hardware remain scarce.

To simulate electronic properties of large-scale materials, two classical obstacles are salient:(i) memory—storing and manipulating a large electronic Hamiltonian can exceed available random-access memory; and (ii) cubic-time diagonalization—extracting electronic properties via full diagonalization of an $N\times N$ matrix scales as $\mathcal{O}(N^3)$\cite{goedecker1999linear}. Because both bottlenecks arise from explicitly storing large matrices and computing their eigenpairs on classical hardware. To bypass full diagonalization, random state (stochastic) strategies were proposed to estimate traces of large sparse matrix by statistical average\cite{hutchinson1990stochastic,iitaka2004random,jin2021randomstate,tang2024random}. Within this paradigm, the time dependent propagation methods (TDPM)\cite{iitaka1997corr,yuan2010tbpm,zhou2023time} and the kernel polynomial method (KPM)\cite{weisse2006kpm} have been successfully applied to large-scale calculation of electronic density of states (DOS)\cite{yuan2010tbpm,le2018electronic}, real-space distributions\cite{iliasov2020hall,yang2020confined,yu2019dodecagonal}, and linear response properties using linear scaling matrix-vector operations for sparse Hamiltonians\cite{andelkovic2018tbgdc,joao2020kite,kuang2021collective,de2023efficient,li2023tbplas,de2024fast}. However, the linear-scale random state methods could also encounter the dilemma of memory storage and computation efficiency as random states and Hamiltonian matrix get huge in simulating large-scale materials. A complementary route is to exploit quantum-computing representations: with $n$ qubits one encodes a $2^{n}$-dimensional Hilbert space  and applies operators as unitary circuits, easing the memory burden\cite{nielsen2010qci,montanaro2016overview}. The required random states are uniform and can be prepared as Haar-random states via low-depth random circuits on quantum computers with a few qubits\cite{harrow2009designs,brandao2016polydesign,schuster2025random}. The random-state approach has recently been realized on quantum hardware and applied to evaluate the density of states in many-body spin systems\cite{wang2023kernel,summer2024dos,goh2024dos}. Building upon these developments, we are motivated to explore random-state quantum algorithms for the computation of electronic properties in realistic materials.

In this work, we reformulate the two random state methods TDPM and KPM, as random state quantum algorithms for electronic-structure property calculations of real materials. Starting from Haar-random states prepared on a qubit register, we apply the electronic real-space Hamiltonian either by real-time propagation, Chebyshev polynomial filtering or a modified phase estimation method to estimate DOS and real-space distribution of electronic eigenstates. We demonstrate the approach using a quantum computer simulator for graphene, fractals, and quasicrystalline bilayer graphene, recovering their characteristic spectral features and spatial patterns. Our findings pave a practical pathway toward scalable electronic-property calculations for large periodic and aperiodic materials on quantum hardware.

\section{METHODS}
\label{sec:methods}

\begin{figure}
    \centering
    \includegraphics[width=1.0\linewidth]{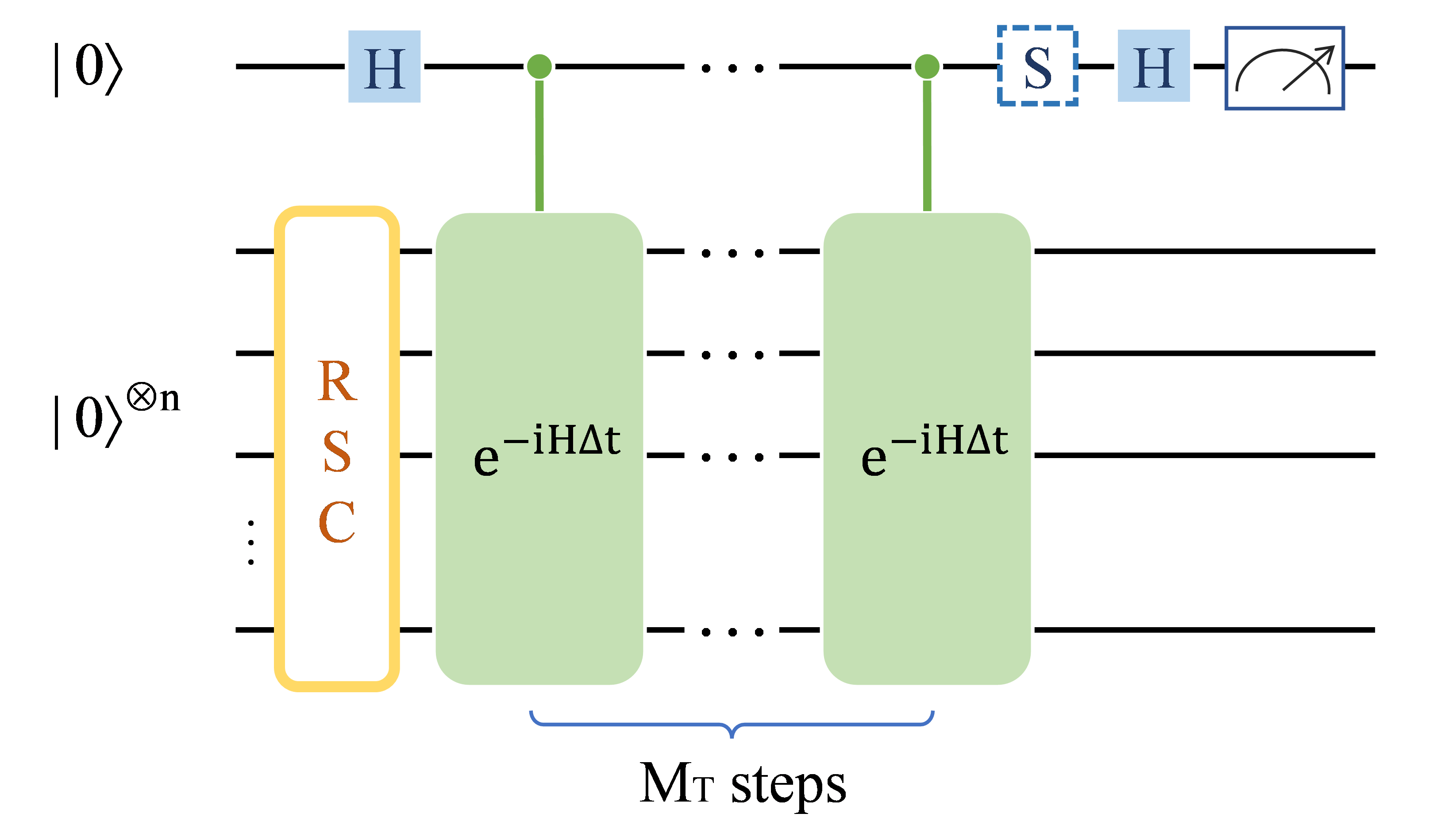}
    \caption{Random state circuit (RSC) for evaluating the time-correlation function \(C^{p}(t)\) for the \(p\)-th stochastic sample using the Hadamard test. The ancilla qubit is initialized in the \(\ket{0}\) state and transformed to the \(\ket{+}\) state via a Hadamard gate. The data register is prepared in a normalized Haar-random state \(\ket{\psi_0^{p}}\), representing the \(p\)-th sample in the ensemble, using the randomized state preparation module (yellow box). Controlled by the ancilla, the time evolution operator \(e^{-iH\Delta t}\) is applied to the data qubits through a Trotterized approximation with a time step $\Delta t$ (green boxes). A second Hadamard gate is applied to the ancilla before measurement. The expectation value of the ancilla yields the real part \(\Re[\bra{\psi_0^{p}}e^{-iHt}\ket{\psi_0^{p}}]\). To access the imaginary part \(\Im[\bra{\psi_0^{p}}e^{-iHt}\ket{\psi_0^{p}}]\), a $S$ gate shown in dotted blue box, is inserted before the final Hadamard.}
    \label{fig:Htest1}
\end{figure}

\begin{figure*}
    \centering
    \includegraphics[width=1.0\textwidth]{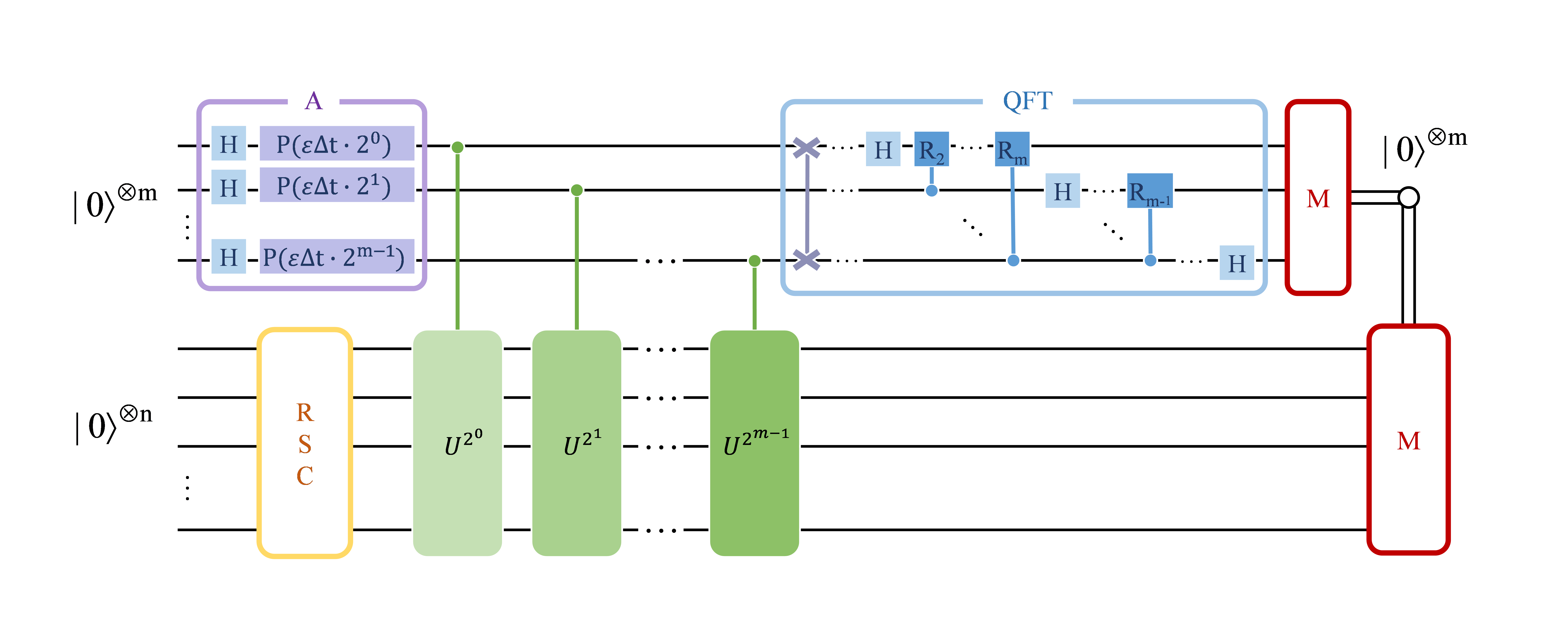}
    \caption{Quantum circuit scheme for calculating quasi-eigenstates based on a modified quantum phase estimation (QPE) protocol combined with RSC. The purple block (labeled with $A$) initializes the ancilla qubits with a sequence of Hadamard and single-ubit controlled-phase gates with the energy level $\varepsilon$ attached. The orange block (denoteds as RSC) prepares the data register in a randomized initial state. The green block applies the set of controlled unitaries $U^j = e^{-iH 2^j \Delta t}$, each conditioned on the $j$th ancilla qubit, while the blue block performs the quantum Fourier transform (QFT) on the ancilla register. Finally, the red blocks represent the measurement stage, where the data qubits are projected only when the ancilla qubits collapse to $\ket{0}^{\otimes m}$, enabling the extraction of quasi-eigenstates at selected energy levels.
    }
    \label{fig:qusiEigen_cir}
\end{figure*}

\subsection{DOS via quantum TDPM}
\label{subsec:method_te}

Electronic density of states plays a pivotal role in the characterization of a material’s electronic structure. It can be calculated with the linear-scaling random-state TDPM ~\cite{yuan2010tbpm,zhou2023time}. In TDPM, the system is initialized in a normalized random superposition of computational basis states:
\begin{equation}
\ket{\psi_{0}^{p}} = \sum_{i} c_{i}^{p}\ket{i}\,,
\label{eq:psi_init}
\end{equation}
where the coefficients satisfy $\sum_{i}\bigl|c_{i}^{p}\bigr|^{2}=1$ for each $p$, with the superscript $p$ indexing the distinct random initial states. For each sample $p$, one can evaluate the time-correlation function
\begin{equation}
C^{p}(t) = \bra{\psi_{0}^{p}} e^{-iHt} \ket{\psi_{0}^{p}}\,,
\label{eq:corr_def_refined}
\end{equation}
with the Plank constant $\hbar = 1$. The DOS is then obtained by ensemble‐averaging over $S$ independent samples\cite{yuan2010tbpm}:
\begin{equation}
D(\varepsilon)= \lim_{S\to\infty}
\frac{1}{S}
\sum_{p=1}^{S}
\frac{1}{2\pi}\int_{-\infty}^{\infty}
e^{i\varepsilon t} C^{p}(t)\mathrm{d}t\,.
\label{eq:dos_final_refined}
\end{equation}
By virtue of ergodicity, this procedure yields statistical convergence to the ideal DOS in the limit of large $S$.

In light of these classical limitations mentioned in Sec.~\ref{sec:intro}, quantum processors have already demonstrated preliminary advantages in random circuit sampling (RCS)~\cite{arute2019quantum,zhong2020quantum,schuster2025random}, 
highlighting their ability to efficiently generate and sample from high-dimensional 
Porter--Thomas–distributed random states beyond the reach of classical methods. Building on this, we combine random state circuits (RSC) with the Hadamard test to evaluate the complex-valued temporal correlation function $C^{p}(t)$ in Eq.(\ref{eq:corr_def_refined}). This quantum routine enables efficient estimation of both the real and imaginary parts of $C^{p}(t)$ at discrete time steps, which are then processed through classical Fourier analysis. The central task of the quantum circuit is to implement the time-evolution operator, starting from the decomposition of a real-space electronic Hamiltonian (see appendix ~\ref{append_sec:TBmodel}) into a sum of Pauli tensor products to facilitate circuit construction.
\begin{equation}
    H = \sum_j c_j P_j,
\end{equation}
where $P_j = \bigotimes_{i=0}^{n-1} \sigma_i$ with each Pauli matrix $ \sigma_i \in \{I, X, Y, Z\}$. This decomposition can be realized via various techniques, including fermion-to-qubit mappings such as the Jordan–Wigner (JW)\cite{jordanwigner1928} and Bravyi–Kitaev (BK) transformations\cite{seeley2012bravyi,bravyi2002fermionic}, as well as linear combination of unitaries (LCU)\cite{childs2012lcu,berry2015taylor}. Time-evolution operator \( e^{-iHt} \) is then implemented via a first-order Trotter decomposition with a finite-step error characterized as $\mathcal{O}(t^2/M_T)$~\cite{trotter1959product}
\begin{equation}
    e^{-iHt} \approx \left( \prod_j e^{-ic_j P_j \Delta t} \right)^{M_T} \,,
\label{eq:trotter_decomp}
\end{equation}
where \( \Delta t = t/M_T \) and \( M_T \) denotes the number of decomposition steps, defining the temporal resolution of the simulation. Each unitary operator \( e^{-ic_j P_j \Delta t} \) is realized as a sequence of basic gates followed by a multi-qubit controlled rotation, more circuit-level details as elaborated in Fig.~\ref{fig:trotter}. 

We adopt the Hadamard test quantum circuit to evaluate the time-correlation function $C^{p}(t)$, as illustrated in Fig.~\ref{fig:Htest1}. The whole circuit consists one ancilla qubit and $n= \log_2 N $ data qubits. The data register can be initialized in a Haar-random state \( \ket{\psi_0^p} \) using RCS strategies\cite{summer2024dos,schuster2025random}. The ancilla qubit is prepared in the \( \ket{+} \) state via a Hadamard gate and used to coherently control the operation of \( U = e^{-iHt} \) on the data register. A subsequent Hadamard gate on the ancilla enables the measurement of the real part of the correlation function \(\Re(\bra{\psi_0^p} U \ket{\psi_0^p}) = 2P_0 - 1\), where \( P_0 \) denotes the probability of measuring the ancilla in state \( \ket{0} \). To access the imaginary component, a phase gate $S$ is applied before the final Hadamard, yielding \( \Im(\bra{\psi_0^p} U \ket{\psi_0^p}) = 2P_0 - 1 \). Combining these two results reconstructs the complex-valued time-correlation function with $U=e^{-iHt}$,
\begin{equation}
\bra{\psi_0^p}U\ket{\psi_0^p}= \Re(\bra{\psi_0^p}U\ket{\psi_0^p}) + i \cdot \Im(\bra{\psi_0^p}U\ket{\psi_0^p})\,.
\label{eq:time_corr_time}
\end{equation}
For each sample $\ket{\psi_{0}^p}$ generated via RCS, the corresponding time-correlation function $C^{p}(t)$ is computed at discrete time intervals. The final DOS is obtained by applying a discrete Fourier transform (DFT) to the ensemble-averaged correlation functions over $S$ statistical samples, as formalized in Eq.~(\ref{eq:dos_final_refined}). 

For an electronic Hamiltonian of dimension \(N\), the above proposed 
quantum-enhanced density-of-states estimation algorithm (Q-TDPM) requires only 
\(1+ \log_2 N \) qubits. 
The random state preparation (RSC) stage can be realized by local random quantum circuits 
of depth \(\operatorname{poly}(\log_2 N)\), which approximate unitary \(t\)-designs and thus 
efficiently generate near–Haar random states~\cite{harrow2009designs,brandao2016polydesign,schuster2025random}. 
To implement the time evolution operator, a first-order Trotter–Suzuki product formula is used. 
Achieving a target simulation accuracy \(\epsilon\) for evolution time \(t\) requires 
\(M_{T}\sim\mathcal{O}\!\bigl(t^{2}/\epsilon\bigr)\) Trotter steps, 
and since each step uses a circuit of depth \(l\) for the Hamiltonian decomposition, 
the total Trotterized circuit depth scales as \(\mathcal{O}\!\bigl(M_{T}\,l\bigr)=\mathcal{O}\!\bigl(t^{2}l/\epsilon\bigr)\)~\cite{emerson2003pseudo,berry2007efficient,trotter1959product,suzuki1976generalized}. 
In addition, the Hadamard test performed over \(S\) independent samples with 
\(N_{\text{shots}}\) measurements per sample incurs a statistical error of 
\(\mathcal{O}\!\bigl(1/\sqrt{S\,N_{\text{shots}}}\bigr)\), implying a total sampling 
cost of \(\mathcal{O}(1/\epsilon^{2})\) to achieve a target precision 
\(\epsilon\) in the final observable estimate~\cite{montanaro2016overview}. 
Combining the contributions from state preparation, Trotterized time evolution 
(depth \(M_{T}l\)) and sampling error, the overall asymptotic complexity of Q-TDPM can be expressed as $\mathcal{O}\!\Biggl[
\frac{1}{\epsilon^{2}}\Bigl(\operatorname{poly}(\log_2 N)+\frac{t^{2}l}{\epsilon}\Bigr)
\Biggr]$, which quantitatively captures the trade-off between circuit depth, simulation 
accuracy, and sampling overhead in the proposed algorithm.




\subsection{DOS via quantum KPM}
\label{subsec:DOS_kpm}

Apart from the time-evolution method described in Sec.~\ref{subsec:method_te}, another widely used classical approach is the KPM approximating the DOS via a Chebyshev polynomial expansion\cite{weisse2006kpm}
\begin{equation}
D(\varepsilon) = \frac{1}{\pi \sqrt{1 - \varepsilon^{2}}} 
\left[ \gamma_{0}^{M} \mu_{0} + 2 \sum_{m=1}^{M} \gamma_{m}^{M} \mu_{m} T_{m}(\varepsilon) \right],
\label{eq:cheby_DOS}
\end{equation}
where $\gamma_m^M$ are Jackson kernel coefficients to suppress Gibbs oscillations, 
$T_m(\varepsilon) = \cos[m \arccos(\varepsilon)]$ $(\varepsilon \in [-1, 1])$ are Chebyshev polynomials, 
and $\mu_m$ are the Chebyshev moments
\begin{equation}
\mu_m = \int_{-1}^{1} D(\varepsilon) T_m(\varepsilon) \, d\varepsilon.
\label{eq:cheby_moments}
\end{equation}
The DOS is defined as $
D(\varepsilon) = \frac{1}{N} \sum_{k=0}^{N-1} \delta(\varepsilon - E_k)$, where $N$ is the Hilbert space dimension of the Hamiltonian $H$ and $\{E_k\}$ are eigenvalues of H. Rescaling $H$ to $
\widetilde{H} = \frac{H-b}{a}$ satisfies the Chebyshev polynomial interval, where scaling parameters $a$ and $b$ correspond to $\frac{E_\text{max}\pm E_{\text{min}}}{2}$ respectively, with $E_{\text{min}}$ and $E_{\text{max}}$ the lower and upper bound of the pristine $H$, that can be estimated via Lanczos tridiagonalization\cite{lanczos1950iteration}.

Inserting the DOS $D(\varepsilon)$ into Eq.~(\ref{eq:cheby_moments}) yields the stochastic trace representation  
\begin{equation}
\begin{aligned}
  \mu_m &= \frac{1}{N} \operatorname{Tr}[T_m(\widetilde{H})] \\ 
      &= \frac{1}{N} \operatorname{Tr}\!\left[\cos\!\left(m \arccos \widetilde{H}\right)\right].  
\end{aligned}
\label{eq:chebymom_trace}
\end{equation}
Using the identity $\arccos(\widetilde{H}) = \frac{\pi}{2} - \arcsin(\widetilde{H})$ and performing a Taylor expansion of $\arcsin(\widetilde{H})$ around the rescaled spectral center $\varepsilon = 0$~\cite{summer2024dos}, Eq.~(\ref{eq:chebymom_trace}) can be generally rewritten in a compact form  
\begin{equation}
\mu_m \approx \cos\frac{m\pi}{2} \, \frac{1}{N} \operatorname{Tr}\!\left[ \cos \left(m H_L \right)\right] 
+ \sin\frac{m\pi}{2} \, \frac{1}{N} \operatorname{Tr}\!\left[ \sin \left(m H_L \right)\right],
\label{eq:cheby_trace2}
\end{equation}
where $H_L$ denotes the truncated odd-power series  
\begin{equation}
H_L = \sum_{l=0}^{L} \frac{(2l)!}{2^{2l}(l!)^2(2l+1)} \, \widetilde{H}^{\,2l+1}.
\end{equation}
According to the parity of $m$, the Chebyshev moments in Eq.~(\ref{eq:cheby_trace2}) can be further simplified as  
\begin{equation}
\mu_m \approx 
\begin{cases}
\dfrac{(-1)^{\frac{m}{2}}}{N} \operatorname{Tr}\!\left[ \cos \left(m H_L \right)\right], & m \ \text{even}, \\[1.0em]
\dfrac{(-1)^{\frac{m-1}{2}}}{N} \operatorname{Tr}\!\left[ \sin \left(m H_L \right)\right], & m \ \text{odd}.
\end{cases}
\label{eq:cheby_parity}
\end{equation}
The quantum subroutine is then used to evaluate $\operatorname{Tr}[\cos(m H_L)]$ and $\operatorname{Tr}[\sin(m H_L)]$ separately, which naturally correspond to the real and imaginary parts of $\operatorname{Tr}(e^{i m H_L})$ via Hadamard test under an ensemble average. Therefore it enables the direct reconstruction of $\mu_m$ for each $m$. The stochastic trace estimator reads
\begin{equation}
\operatorname{Tr}(e^{i m H_L})
\;\approx\; \frac{N}{S}\sum_{p=1}^{S}
\bra{\psi_0^{p}}\, e^{i m H_L}\, \ket{\psi_0^{p}}\,.
\label{eq:stoch-trace}
\end{equation}
For each random initial state $\ket{\psi_0^{p}}$ sampled via RCS, the overlap 
$\bra{\psi_0^{p}} e^{i m H_L} \ket{\psi_0^{p}}$ is evaluated with the circuit in Fig.~\ref{fig:Htest1}, where 
$U(m)=e^{i m H_L}$ is implemented by a Trotterized approximation of $e^{i H_L}$ with a fixed number of steps per operation. 
Ensemble averaging over $S$ samples suppresses statistical fluctuations and yields the unbiased stochastic–trace estimator in 
Eq.~(\ref{eq:stoch-trace}). Separate Hadamard tests are employed to access the real and imaginary parts of the trace, yielding 
$\operatorname{Tr}[\cos(m H_L)]$ for even $m$ and $\operatorname{Tr}[\sin(m H_L)]$ for odd $m$, 
and thereby providing the Chebyshev moments $\mu_m$ for each order $m$. 
By truncating the effective Hamiltonian $H_L$ at order $L$ (e.g. $L=2$ for the cases demonstrated in Sec.~\ref{sec:application}),
the Chebyshev moments up to order $M$ can be evaluated and subsequently substituted into 
Eq.~(\ref{eq:cheby_DOS}) to reconstruct the density of states $D(\varepsilon)$ 
at the rescaled energy $\varepsilon$.

For an electronic Hamiltonian of dimension \(N\), the KPM quantum algorithm (Q-KPM) requires the same number of qubits as Q-TDPM, namely \(1+\log_2 N\), and likewise employs random-state preparation circuits of depth \(\mathcal{O}(\log_2 N)\) to approximate unitary \(t\)-designs. To reach a target precision \(\epsilon\), the Chebyshev expansion is truncated at order \(M=\mathcal{O}\!\bigl(\log_2(1/\epsilon)\bigr)\)~\cite{Weiße2006,berry2015taylor}. Because \(H_{L}\approx (H-b\mathbb{I})/a\), the evolutionary operator \(e^{iM H_{L}}\) satisfies \(e^{iM H_{L}}\approx e^{i(M/a)H}e^{-i(Mb/a)}\) of which the second factor $e^{-i(Mb/a)}$ is a global phase and can be ignored, so simulating \(e^{iM H_{L}}\) is equivalent to simulating \(e^{i(M/a)H}\) with total effective time \(t=M/a\). This evolution is numerically implemented by a two-level Trotterisation: (i) decompose \(e^{iM H_{L}}\) into \(M\) unit-time segments \((e^{iH_{L}})^{M}\); (ii) approximate each segment with a first-order Trotter–Suzuki formula using \(n_{T}\) substeps of size \(\delta\tau=1/n_{T}\), such that
$e^{iH_{L}}\approx\Bigl[\prod_{j=1}^{l}e^{iH_{L,j}\delta\tau}\Bigr]^{n_{T}}$ with $H_{L}=\sum_{j=1}^{l}H_{L,j}$.
For first order the error per segment is \(\mathcal{O}(1/n_{T})\)~\cite{trotter1959product}, so distributing a total Trotter error budget \(\epsilon\) evenly across all \(M\) segments gives \(n_{T}=\mathcal{O}(M/\epsilon)\) and a total of substeps \(M n_{T}=\mathcal{O}(M^{2}/\epsilon)\). If each substep requires circuit depth \(l\) due to Hamiltonian decomposition, the overall Trotterised depth scales as $\text{depth}_{\text{trot}}=\mathcal{O}\!\Bigl(\frac{M^{2}}{\epsilon}\,l\Bigr)$ in the \(H_{L}\) unit, or expressed in physical time
$\text{depth}_{\text{trot}}=\mathcal{O}\!\Bigl(\frac{t^{2}}{\epsilon}\,l\Bigr)
=\mathcal{O}\!\Bigl(\frac{M^{2}}{a^{2}\epsilon}\,l\Bigr)$.
Thus segmenting \(e^{iM H_{L}}\) into \(M\) pieces and Trotterising each piece yields the same asymptotic depth scaling as simulating \(e^{i(M/a)H}\) directly, up to constant factors. Together with the \(\operatorname{poly}(\log_2 N)\) depth for random-state preparation and the \(\mathcal{O}(1/\epsilon^{2})\) sampling cost of the Hadamard test~\cite{montanaro2016overview}, this gives the full resource estimate for Q-KPM as $\mathcal{O}\!\Biggl[
\frac{1}{\epsilon^{2}}\Bigl(\operatorname{poly}(\log_2 N)+\frac{M^{2}l}{a^2\epsilon}\Bigr)
\Biggr]$.
\begin{figure*}
    \centering
    \includegraphics[width=1\linewidth]{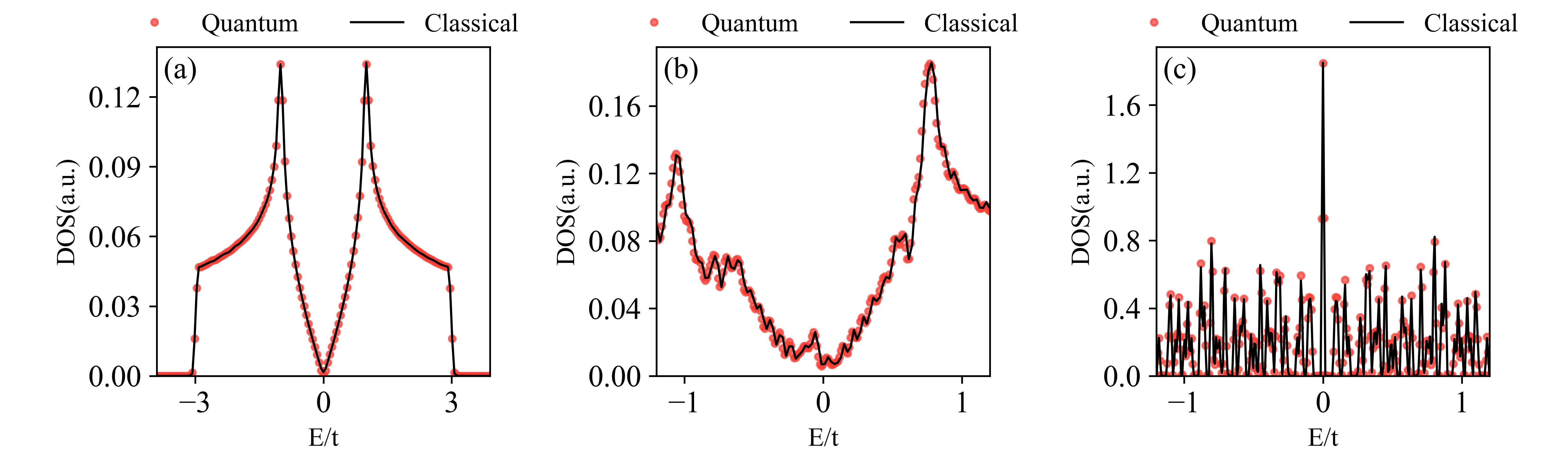}
    \caption{Comparison of Q-TDPM (solid black line) and classical numerical (red circles) density of states (DOS) calculated via time evolution with 1000 random samples. (a) DOS for a 64×64 graphene lattice. (b)DOS for a 4 nm radius disk of 30°-twisted bilayer graphene (30°-tBG, 3828 atoms) . (c) DOS for a Sierpiński carpet fractal (256 atoms).}
    \label{fig:DOS_te}
\end{figure*}
\begin{figure*}
    \centering
    \includegraphics[width=1\linewidth]{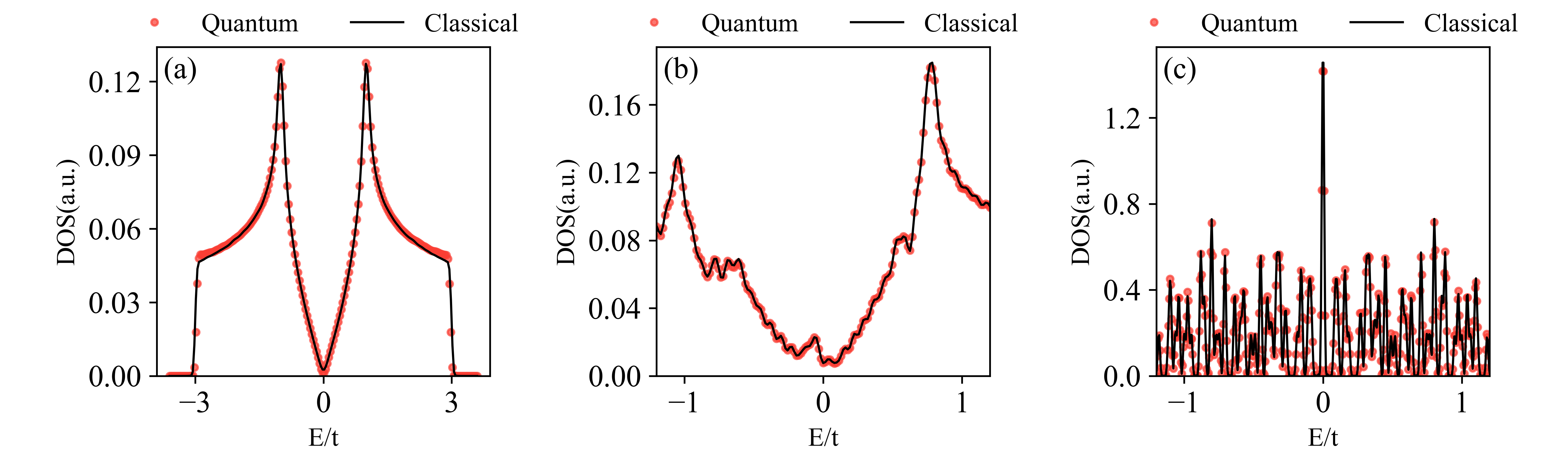}
    \caption{Comparison of Q-KPM (solid black line) and classical numerical (red circles) density of states (DOS) calculated via the kernel polynomial method (KPM) with 1000 random samples. (a) DOS for a 64×64 graphene lattice. (b)DOS for a 4 nm radius disk of 30°-twisted bilayer graphene (30°-tBG, 3828 atoms) . (c) DOS for a Sierpiński carpet fractal (256 atoms).}
    \label{fig:DOS_kpm}
\end{figure*}

\subsection{Spatial Distribution of Electronic States}
\label{subsec:quasieigen}

Spatial distribution of electronic eigenstates is another fundamental property of materials. 
It is characterized by the local density of states (LDOS) $\rho(\mathbf r,\varepsilon)$, 
which encodes the spatially resolved density of electronic states per unit energy at position 
$\mathbf r$ and energy $\varepsilon$~\cite{tersoff1985stm}. 
Inspired by the time-evolution approach based on randomized states proposed in Ref.~\cite{rall2020quantum}, 
we instead approximate spectral properties within a quantum-computational framework.
In the following, we propose a modified quantum phase estimation (M-QPE) method that incorporates the RCS 
strategy to directly evaluate spatial distribution of quasi-eigenstates\cite{yuan2010tbpm} at selected energy levels.
In the framework, the construction of quasi-eigenstates begins with a randomized initial state 
$\ket{\psi^p_0} = \sum_{r_i} C_{r_i} \ket{r_i}$, where $\{\ket{r_i}\}$ represent local lattice-site wave functions that can be encoded in the qubit-indexed basis. The subsequent time evolution $\ket{\psi^p(t)} = e^{-iHt}\ket{\psi^p_0}$ is Fourier transformed to extract the spectral features of the Hamiltonian, yielding
\begin{equation}
\begin{aligned}
\ket{\Psi^p(\varepsilon)} =& \frac{1}{2\pi} \int_{-\infty}^{\infty} e^{i\varepsilon t} \ket{\psi^p(t)} dt \\
=& \frac{1}{2\pi} \int_{-\infty}^{\infty} e^{i(\varepsilon -H) t} \ket{\psi^p_0} dt\\
=& \sum_{r_i} C_{r_i} \delta(\varepsilon - E_{i}) \ket{r_i}\,.  
\end{aligned}
\label{eq:qusieigen_define}
\end{equation}
The final quasi-eigenstate $\ket{\Psi(\varepsilon)}$ is then obtained by averaging over all random states, thereby defining the energy-resolved quasi-eigenstates  of the electronic Hamiltonian at the energy level $\varepsilon$~\cite{yuan2010tbpm}. Since the quasi-eigenstate wavefunction can be expanded in the local indexed basis $\{\ket{r_i}\}$, taking the squared modulus of the following projections yields quasi-eigenstates at a given energy,
\begin{equation}
\rho(\mathbf{r}_i, \varepsilon) = \left| \langle r_i | \Psi(\varepsilon) \rangle \right|^2\,,
\label{eq:ldos_def}
\end{equation}
which characterizes the real-space probability distribution of the electronic state localized at site $\mathbf{r}_i$ with energy $\varepsilon$. Equivalently, the quasi-eigenstate can be expressed by discretizing the Fourier integral into $M$ temporal steps with $\Delta t = t/M$:
\begin{equation}
\ket{\Psi^p(\varepsilon)} = \frac{1}{\sqrt{M}} \sum_{k=0}^{M-1} 
e^{i\varepsilon k\Delta t} \ket{\psi^p(k\Delta t)}, 
\quad k \in \mathbb{Z}\,.
\label{eq:qusieigen_QFT}
\end{equation}

In the following, we demonstrate how the wavefunction in Eq.~(\ref{eq:qusieigen_QFT}) can be realized within a M-QPE framework. As shown in Fig.~\ref{fig:qusiEigen_cir}, the quantum circuit is composed of $m$ ancilla qubits and $n$ data qubits, 
where $m = \log_2 M$ ensures that the $M$ discrete time steps are represented in binary form and $n=\log_2 N$ with $N$ the dimensionality of lattice Hamiltonian. 
The ancilla register is first prepared by applying a layer of Hadamard gates, followed by position-dependent phase gates, 
such that the $j$th ancilla qubit acquires a phase $P(\varepsilon \Delta t \cdot 2^j)$. 
Through this procedure, the ancilla register is prepared in the superposition state indicated by block~$A$:
\begin{equation}
\begin{aligned}
 \ket{\psi_A} =&\prod_{j=0}^{m-1} P\!\left(\varepsilon \Delta t \cdot 2^j\right) H^{\otimes m} \ket{0}^{\otimes m} \\
 =& \frac{1}{\sqrt{M}}\Bigg(\prod_{j=0}^{m-1} e^{\,i(\varepsilon \Delta t\,2^j)k_j}\Bigg)\ket{k}  \\
 =&\frac{1}{\sqrt{M}} \sum_{k=0}^{M-1} e^{\,i\varepsilon \Delta t \sum_{j=0}^{m-1} 2^j k_j}\ket{k} \\
 =&\frac{1}{\sqrt{M}} \sum_{k=0}^{M-1} e^{\,i \varepsilon k \Delta t}\ket{k} \,.
\end{aligned}
\label{eq:psiA_init}
\end{equation}

The data register is prepared in a Haar-random state $\ket{\psi^p_0}$ via RCS, while the ancilla register is initialized in the state $\ket{\psi_A}$. 
For each ancilla qubit $j$, we define the controlled unitary 
\begin{equation}
 U^{j} = \ket{0}\!\bra{0}_j \otimes \mathbb{I} \;+\; \ket{1}\!\bra{1}_j \otimes e^{-iH\,2^j\Delta t} \,. 
\end{equation}
Let $\mathcal{U} \equiv \prod_{j=0}^{m-1} U^{(j)}$. Acting on a computational basis state $\ket{k}=\bigotimes_{j}\ket{k_j}$, we obtain
\begin{equation}
\begin{aligned}
\mathcal{U}\big(\ket{\psi^p_0}\otimes\ket{k}\big)
&=\prod_{j=0}^{m-1}\Big(e^{-iH\,2^j\Delta t}\Big)^{k_j}\ket{\psi^p_0}\otimes\ket{k} \\[4pt]
&=e^{-iH\left(\sum_{j=0}^{m-1}2^j k_j\right)\Delta t}\ket{\psi^p_0}\otimes\ket{k} \\[4pt]
&= e^{-iH k \Delta t}\ket{\psi^p_0}\otimes\ket{k},
\end{aligned}
\label{eq:ctrlU_psi}
\end{equation}
where $k=\sum_{j=0}^{m-1} 2^j k_j$. Defining the time-evolved data state as $\ket{\psi^p(k\Delta t)}\equiv e^{-iHk\Delta t}\ket{\psi^p_0}$, 
and applying $\mathcal{U}$ to the ancilla superposition, we obtain the entangled ancilla–data state
\begin{equation}
\begin{aligned}
\ket{\psi_{A+D}}
&=\mathcal{U}\Big(\ket{\psi^p_0}\otimes\ket{\psi_A}\Big) \\[4pt]
&= \frac{1}{\sqrt{M}} \sum_{k=0}^{M-1} e^{i\varepsilon k\Delta t}\,\ket{\psi^p(k\Delta t)}\otimes\ket{k}.
\end{aligned}
\label{eq:entangled_state_detail}
\end{equation}

A quantum Fourier transform (QFT) is then applied to the ancilla register, producing
\begin{equation}
\ket{\psi_{A+D}'} = \frac{1}{M} \sum_{j=0}^{M-1} 
\left( \sum_{k=0}^{M-1} e^{i\varepsilon k\Delta t} e^{2\pi i jk / M} \ket{\psi^p(k\Delta t)} \right) \otimes \ket{j}.
\end{equation}

By post-selecting the ancilla register on the outcome $j=0$, 
the data register collapses to
\begin{equation}
\ket{\psi_D} = \frac{1}{\sqrt{M}} \sum_{k=0}^{M-1} e^{i\varepsilon k\Delta t}\,\ket{\psi^p(k\Delta t)}
\;\equiv\; \ket{\Psi^p(\varepsilon)}\,,
\end{equation}
which corresponds to a quasi-eigenstate associated with energy $\varepsilon$. Subsequent measurements of the data qubits in the computational basis map $\ket{\Psi(\varepsilon)}$ onto the local lattice basis ${\ket{\mathbf{r}_i}}$, resulting in the real-space probability density distribution of the targeted energy as defined in Eq.~(\ref{eq:ldos_def}). To enhance the success probability of projecting the ancilla register onto the state $\ket{j=0}$, one can apply Grover-type amplitude amplification operators~\cite{grover1997search,brassard2002quantum}: by marking $\ket{j=0}$ as the target subspace and iteratively applying the Grover diffusion operator, the amplitude associated with $\ket{j=0}$ can be quadratically boosted, thereby improving the overall efficiency of the algorithm. In the application section, we adopt post-selection on the ancilla register to validate the effectiveness of the proposed algorithm.

The M-QPE algorithm based on RSC for spatial DOS estimation requires \(\log_2 M\) ancilla qubits in addition to the system register \(\log_2 N \), and its total complexity is contributed by several parts. 
First,  RSC approximates Haar–random states using pseudo–random circuits of depth \(\mathrm{poly}(\log_2 N)\)~\cite{emerson2003pseudo}, more recently leading to a state–preparation overhead \(\mathcal{O}(\log_2 N)\)~\cite{schuster2025random}. 
Second, each time step \(\ket{\psi^p(k\Delta t)}\) involves a controlled evolution \(e^{-iH\Delta t}\), which in the QPE procedure appears as controlled $e^{-iH\,2^{m+1}\Delta t}$ with decomposition depth \(l\); for first–order Trotter–Suzuki one can choose 
\(M_{T}=\mathcal{O}(t^{2}/\epsilon)\) with \(t=2^{m+1}\delta t\) to keep the Trotter error below \(\epsilon\), giving a Hamiltonian–simulation cost \(\mathcal{O}(t^{2}l/\epsilon)\)~\cite{trotter1959product,suzuki1976generalized}. 
Third, the quantum Fourier transform acting on \(\log_2 M\) ancilla qubits requires \(\mathcal{O}(\log^2_2 M)\) gates. 
Fourth, post–selection on the ancilla outcome \(j=0\) succeeds with probability \(P=\mathcal{O}(1/M)\) and is boosted to \(\mathcal{O}(1)\) using amplitude amplification with overhead \(\mathcal{O}(1/\sqrt{P})=\mathcal{O}(\sqrt{M})\)~\cite{brassard2002quantum}. 
Finally, estimating observables over \(S\) random states with \(N_{\mathrm{shots}}\) circuit repetitions incurs a statistical error \(\mathcal{O}(1/\sqrt{S\,N_{\mathrm{shots}}})\), implying a sampling cost \(\mathcal{O}(1/\epsilon^{2})\) to achieve precision \(\epsilon\). 
Combining these contributions, the overall asymptotic resource requirement of the algorithm can be summarized as \(\mathcal{O}\!\bigl[\frac{1}{\epsilon^{2}}\bigl(\log_2 N + \frac{t^{2}l}{\epsilon} + \log_2^2 M + \sqrt{M}\bigr)\bigr]\) with \(t=2^{m+1}\Delta t\), which explicitly shows the dependence on the target precision \(\epsilon\) through the Trotter step size and the sampling overhead.
For the typical setting \(M\sim N\) and sparse Hamiltonians with \(l=\mathcal{O}(\log_2 N)\)~\cite{berry2007efficient,childs2012lcu}, the dominant cost arises from the controlled time evolution term 
\(\frac{t^{2}l}{\epsilon}\). After multiplying by the sampling factor \(\frac{1}{\epsilon^{2}}\) this yields a leading scaling of 
\(\mathcal{O}(N^{2}\Delta^2 t\log_2 N/\epsilon^{3})\) for the overall complexity.


\section{Application}
\label{sec:application}

\begin{figure}
    \centering
    \includegraphics[width=1.0\linewidth]{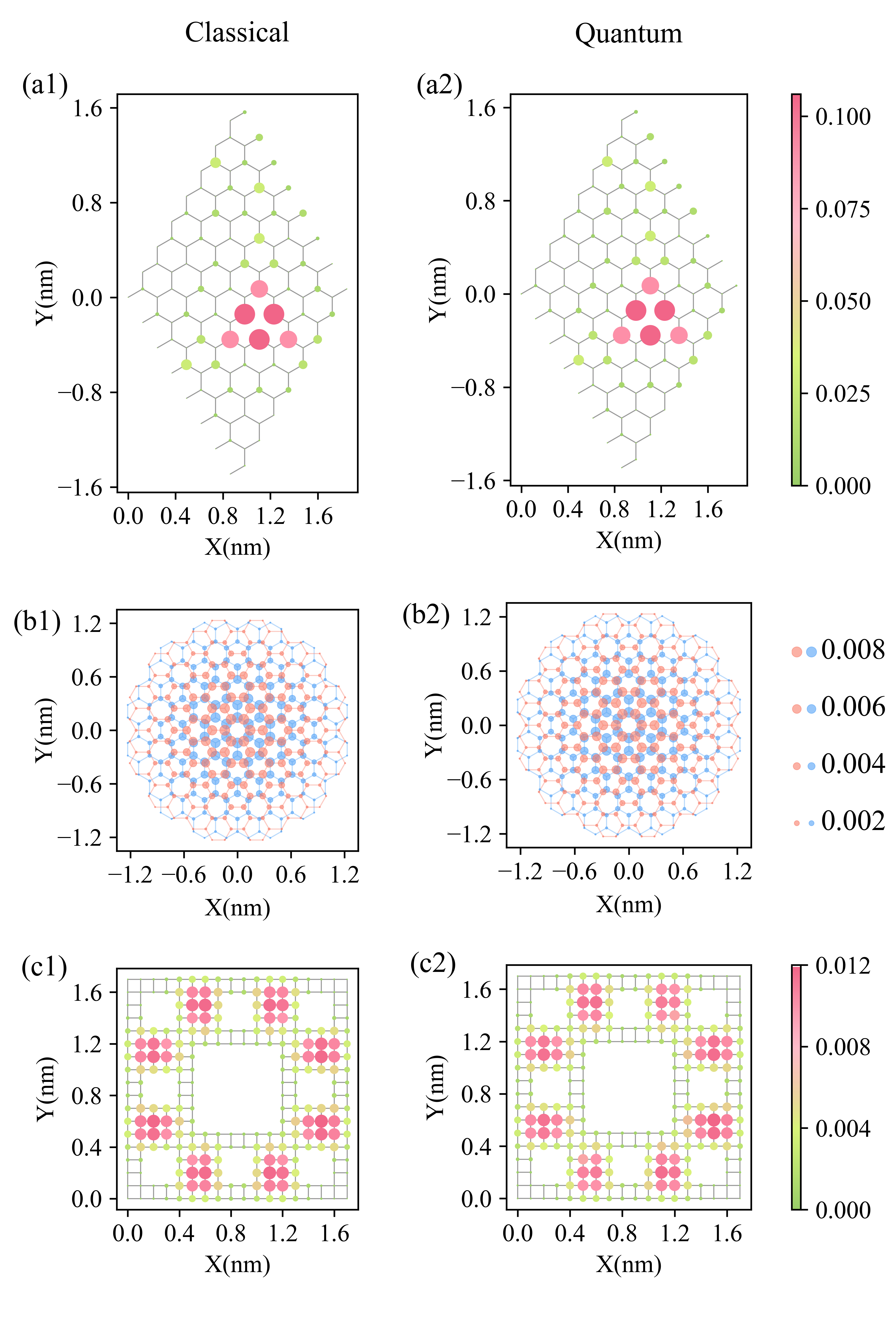}
    \caption{Comparison of quantum (left) and classical (right) quasi-eigenstates maps with 300 random samples. (a) Quasi-eigenstate maps at \(0\ eV\) for an 8×8 graphene lattice with a single vacancy (127 atoms). (b) Quasi-eigenstate maps at minimum eigenvalue \(-11.39\ eV\) for a 1.3 \(nm\) radius disk of \(30^\circ\)-twisted bilayer graphene (408 atoms). (c) Quasi-eigenstate maps at \(-3.41\ eV\) for a Sierpiński carpet fractal (256 atoms).}
    \label{fig:eigen}
\end{figure}

To validate the random-state quantum algorithms proposed in Sec.~\ref{sec:methods} for calculating DOS and quasi-eigenstates, we benchmark three representative two-dimensional systems covering both periodic and aperiodic structures, namely monolayer graphene~\cite{neto2009graphene}, twisted bilayer graphene quasicrystal~\cite{ahn2018dodecagonal}, and the Sierpiński carpet fractal lattice~\cite{vanveen2016sierpinski}. 
All quantum algorithms based on random states were implemented using the Qiskit framework and executed on the \textit{``statevector''} simulator backend without sampling noise~\cite{javadi2024qiskit}. Random quantum states were generated via \texttt{qiskit\allowbreak.quantum\_info\allowbreak.random\_statevector}. The system Hamiltonians were decomposed into sums of Pauli strings using \texttt{SparsePauliOp\allowbreak.from\_operator}, and time-evolution operators were constructed based on the first-order Trotter–Suzuki decomposition (see Appendix~\ref{append_sec:eiHt_to_cir}), which were subsequently compiled into quantum circuits. To validate the accuracy of the quantum simulations, results were benchmarked against classical random-state methods (see Appendices~\ref{append_sec:DOS_kpm_classical} and~\ref{append_sec:quasi_eig_app}).

For periodic pristine graphene, the DOS is evaluated on a $64 \times 64$ supercell (8192 atoms) using a nearest-neighbor tight-binding (TB) Hamiltonian (see appendix ~\ref{append_sec:TBmodel}). In Q-TDPM, the simulation employs a time step of $1/48$ and $M_T =1000$ steps, whereas in Q-KPM we employ a 250-term Chebyshev expansion combined with a 4-term Trotter decomposition for exponential evaluation $e^{iH_L}$ with approximation order $L=2$. As shown in Fig. \ref{fig:DOS_te} (a) and \ref{fig:DOS_kpm} (a), both Q-TDPM and Q-KPM accurately reproduce the Dirac-point DOS minimum at $E=0$ as well as the van Hove singularities(VHS) at the $M$ points near $E=\pm t$ with the standard graphene hopping $t\approx2.7\,$eV~\cite{neto2009graphene,reich2002tight}. To further validate real-space distribution of electronic states, we consider an $8 \times 8$ graphene sheet containing a single vacancy, evolve with time step of $0.1$ for $256$ steps, and compuate the M-QPE based LDOS maps. The exponential operator $e^{-iH\Delta t}$ is approximated using a 5-term Trotter decomposition to enhance the accuracy of the time evolution. The resulting LDOS clearly exhibits vacancy-induced localized states with 3-fold symmetry around the defect, consistent with classical TB results (see Fig.\ref{fig:eigen} (a)) \cite{pereira2006disorder,li2023tbplas}.

As a quasicrystal, twisted bilayer graphene at $30^\circ$ (30$^\circ$-TBG) exhibits long-range 12-fold rotational symmetry but lacks translational order, posing challenges to conventional electronic-structure calculations~\cite{ahn2018dodecagonal}. 
To validate the DOS and quasi-eigenstate qauntum algorithms on quasicyrstal samples, we examine finite round flakes: a disk of radius 4\,nm containing 3828 atoms for DOS calculations (see Fig.~\ref{fig:atomicstructure}) and a smaller disk of radius 1.3\,nm with 408 atoms for quasi-eigenstates. 
In Q-TDPM, the DOS is simulated with a time step $\Delta t=1/48$ for 2400 steps, whereas in Q-KPM we adopt a 600-term Chebyshev expansion with 4 Trotter steps for implementing the exponential operator $e^{iH_L}$ with approximation order $L=2$. 
As shown in Fig.~\ref{fig:DOS_te}(b) and Fig.~\ref{fig:DOS_kpm}(b), the resulting DOS captures the additional VHS characteristic of 30$^\circ$-TBG~\cite{yu2019dodecagonal}, in agreement with classical benchmarks under the same parameter settings.
Because the limited memory of the current quantum simulator restricts the number of qubits---and thus confines our quantum simulations to a 4\,nm disk~\cite{yu2019dodecagonal} where the Dirac cone is suppressed---we increased the disk radius to 20\,nm in classical TB calculations, where the Dirac cone reappears (see Appendix~\ref{append_sec:DOS_te_app}), confirming that its absence in the quantum results is a finite-size effect.
For the calculation of quasi-eigenstates part, the smaller disk is evolved with a time step of 0.1 for 512 steps,where the time-evolution operator is evaluated via a 5-term Trotter decomposition. The resulting quasi-eigenstate maps shown in Fig.~\ref{fig:qusiEigen_cir}(b) clearly display the expected 12-fold (dodecagonal) rotational symmetry with quasi-localized patterns, consistent with previous studies~\cite{moon2019quasicrystalline,yu2019dodecagonal}.

Next, we benchmark the random state quantum algorithms on a finite Sierpi\'nski carpet fractal.  
The lattice is generated by iteratively removing the central block from each \(3\times3\) plaquette; the iteration index \(I\) sets the fractal order and geometric complexity~\cite{vanveen2016sierpinski,yang2020confined}.  
For the quantum simulations we take \(I=2\), corresponding to 256 atoms in our discretization.  
In the DOS calculations, Q-TDPM uses a time step \(\Delta t=1/128\) for 48{,}000 steps, whereas Q-KPM employs a 1500-term Chebyshev expansion with a 32-step internal Trotter decomposition for the time-evolution operator.  
The resulting DOS, shown in Fig.~\ref{fig:DOS_te}(c) and Fig.~\ref{fig:DOS_kpm}(c), reproduces the expected particle--hole symmetry and the discrete spectral features characteristic of finite Sierpi\'nski carpets~\cite{yang2020confined,yao2023energy}, in agreement with our classical benchmarks and previous studies\cite{vanveen2016sierpinski,yang2020confined}.  
For quasi-eigenstates, a 256-atom lattice sample is evolved with \(\Delta t=0.1\) for 256 steps using M-QPE, employing again a 10-term Trotter decomposition for $e^{-iH\Delta t}$.  
The resulting maps reveal self-similar, hierarchical spatial patterns that mirror the underlying fractal geometry~\cite{vanveen2016sierpinski,yang2020confined}, again in good agreement with classical reference data as shown in Fig.~\ref{fig:eigen}(c).

These numerical benchmarks demonstrate that the random-state quantum algorithms---Q-TDPM, Q-KPM, and M-QPE---can effectively compute electronic-structure observables for both periodic and aperiodic systems, provided that their real-space tight-binding (TB) Hamiltonians are available.
Tables~\ref{tab:qc_resources} and~\ref{tab:QTDPM_QKPM} summarize the corresponding quantum resource requirements and simulation parameters for three representative systems: graphene, 30$^\circ$-twisted bilayer graphene (TBG), and fractal lattices. 
As indicated, the number of qubits required grows only logarithmically with the Hamiltonian dimension \(N\), confirming the exponential encoding efficiency of quantum representations discussed in Sec.~\ref{sec:methods}. For example, as shown in Table~\ref{tab:qc_resources}, simulating a graphene lattice with \(N = 8192\) atoms requires only 14 qubits (including ancillary qubits) and a circuit depth of \(l = 1.46\times10^{6}\) for a single controlled \(e^{-iH\Delta t}\) operation. 
By contrast, the 30$^\circ$-TBG system with \(N = 3828\) atoms exhibits a much higher depth of \(l = 3.95\times10^{8}\), owing to its quasi-periodic interlayer coupling network, while the total qubit requirement increases only slightly to 13. 
For the smaller fractal lattice with \(N = 256\), only 9 qubits are needed, and the circuit depth is reduced to \(2.8\times10^{4}\), highlighting the scalability advantage in simpler lattice geometries. As summarized in Table~\ref{tab:QTDPM_QKPM}, the total depth in the DOS estimation is dominated by the Trotterized time-evolution segments. 
Consistent with the theoretical resource analysis, the overall circuit depth scales as \(\mathcal{O}(M_{T}l)\) for Q-TDPM, where \(M_{T} = t / \Delta t\), and as \(\mathcal{O}(Mn_{T}l)\) for Q-KPM, where \(M\) and \(n_{T}\) represent the Chebyshev expansion order and the number of Trotter substeps, respectively. 
Taking the graphene system as a reference, both approaches exhibit comparable total evolution depths on the order of \(10^{9}\). 
Specifically, for Q-TDPM with \(\Delta t = 1/48\) and \(M_T = 1000\), and for Q-KPM with parameters \(a=12\), \(n_T=4\), and \(M=250\), the total controlled time-evolution depth is of the same magnitude, confirming that both methods achieve equivalent computational scaling in practice. 
This equivalence arises because the Q-KPM decomposition effectively redistributes the time-evolution segments into Chebyshev-ordered substeps without altering the overall scaling factor. 
These quantitative analyses highlight that, while large-scale Hamiltonian simulations remain gate-depth intensive, the logarithmic growth of qubit resources and the modular, reusable time-evolution structure make random state quantum algorithms highly promising for near-term implementation on scalable quantum hardware.
\begin{table}[htbp]
\centering
\fontsize{8}{12}\selectfont
\caption{Quantum resource estimation for the controlled \( e^{-iH\Delta t} \) operation in three representative systems. 
The parameter \( N \) indicates both the Hamiltonian dimension and the number of atoms in the system, with qubit counts including ancillary qubits. 
Circuit depths \( l \) considering the decomposition of multi-qubit gates (in units of \( 10^{3} \)) represent a single controlled evolution step.
}
\label{tab:qc_resources}
\begin{tabular}{%
>{\centering\arraybackslash}p{2cm}  
>{\centering\arraybackslash}p{1.5cm}
>{\centering\arraybackslash}p{1.5cm}
>{\centering\arraybackslash}p{1.5cm}  
}
\toprule
\textbf{System} & \textbf{$N$} & \textbf{Qubits} & \textbf{$l$ ($10^{3}$)} \\
\midrule
\multirow{3}{*}{Graphene} & 512   & 10 & 57    \\
         & 2048  & 12 & 297   \\
         & 8192  & 14 & 1459  \\
\midrule
\multirow{3}{*}{30$^\circ$-TBG}  & 408   & 10 & 5039  \\
         & 948   & 11 & 16987 \\
         & 3828  & 13 & 394704\\
\midrule
\multirow{3}{*}{Fractal}  & 256   & 9  & 28    \\
         & 2048  & 14 & 773   \\
         & 16384 & 15 & 24623 \\
\bottomrule
\end{tabular}
\end{table}
\begin{table}[htbp]
\centering
\fontsize{8}{12}\selectfont
\caption{Parameters used for DOS calculation in different systems using Q-TDPM and Q-KPM methods. 
For Q-TDPM, a controlled \(e^{-iH\Delta t}\) is applied \(M_T\) times. 
For Q-KPM, a controlled \(e^{iH_L \delta t}\) is applied \(Mn_T\) times, where \(\delta t = 1/(a n_T)\), \(a\) is the rescaling factor of the Hamiltonian \(H\), and \(n_T\) is the number of Trotter substeps per segment. 
To ensure consistent calculation accuracy, \(\Delta t = \delta t/a \) and \(M_T = n_T M\), such that the number of controlled \(e^{-iH\Delta t}\) and controlled \(e^{iH_L \delta t}\) operations in the circuit are equal.}
\label{tab:QTDPM_QKPM}
\begin{tabular}{
    >{\centering\arraybackslash}p{1.5cm} 
    >{\centering\arraybackslash}p{1cm} 
    >{\centering\arraybackslash}p{1.1cm} 
    >{\centering\arraybackslash}p{1.2cm} 
    >{\centering\arraybackslash}p{0.8cm} 
    >{\centering\arraybackslash}p{0.8cm} 
    >{\centering\arraybackslash}p{1cm} 
}
\toprule
\multirow{2}{*}{\textbf{System}} & 
\multirow{2}{*}{\textbf{N}} &
\multicolumn{2}{c}{\textbf{Q-TDPM}} &
\multicolumn{3}{c}{\textbf{Q-KPM}} \\
\cmidrule(lr){3-4} \cmidrule(lr){5-7}
& & \(\Delta t\) & \(M_T\) & \(a\) & \(n_T\) & \(M\) \\
\midrule
Graphene & 8192  & 1/48  & 1000   & 12  & 4   & 250   \\
30$^\circ$-TBG  & 3828  & 1/48  & 2400   & 12  & 4   & 600   \\
Fractal  & 256   & 1/128 & 48000  & 4   & 32  & 1500  \\
\bottomrule
\end{tabular}
\end{table}

\section{Conclusion and Discussion}
\label{sec:summary}

We have developed three random-state quantum algorithms—Q-TDPM, Q-KPM and M-QPE—for calculating electronic-structure properties.
Electronic DOS is evaluated with Q-TDPM and Q-KPM through a pipeline comprising random-state circuit preparation, Hamiltonian and Trotter decomposition, Hadamard tests via ancilla qubits, and statistical averaging over multiple random-state circuits.
The proposed random-state M-QPE algorithm obtains the spatial distribution of electronic states by preparing random states, performing Hamiltonian and Trotter decomposition, applying QFT, and using post-selection and direct quantum measurements with multiple ancilla qubits. Moreover, we have analyzed the circuit complexity of all three algorithms and shown that the dominant cost arises from the Trotter decomposition.
Then, their validity has been demonstrated by simulating periodic and quasi-periodic graphene-based materials and non-periodic finite systems and benchmarking the results against their corresponding classical random-state algorithms.

The required quantum resources for simulating these systems have been analyzed and found to be consistent with our theoretical complexity estimates. For finite fractal systems, the number of qubits required is determined by the system size, while statistical convergence can be achieved by increasing the number of independent random-state circuits. For infinite periodic systems such as graphene and twisted bilayer graphene (TBG) quasicrystals, valid results can still be obtained using only a few qubits by constructing sufficiently large finite samples and employing appropriate random circuit sampling. Therefore, random-state quantum algorithms are expected to be particularly beneficial for simulating quasi-infinite materials with large-scale supercells—such as moir\'e  materials~\cite{carr2020electronic,liu2025dpmoire}—on near-term noisy intermediate-scale quantum (NISQ) devices through circuit-level parallelization over random states. On the other hand, the circuit depth is mainly determined by the Trotter decomposition, where a small time step and sufficient number of evolution steps are required to achieve reasonable precision in electronic-structure calculations. Consequently, the practical implementation of random-state quantum algorithms on current NISQ hardware may still be constrained by noise accumulation in deep circuits. The impact of quantum noise on the accuracy and scalability of these algorithms will be the focus of our future investigations.

Here, the random-state quantum algorithms are primarily designed to compute fundamental electronic-structure properties of materials. Beyond these applications, they can be extended to evaluate electron densities~\cite{zhou2023time}, electronic transport, and optical responses~\cite{li2023tbplas} by incorporating additional operator decompositions into the quantum circuits, which will be explored in future work. Recent progress in random-state circuit (RSC) preparation has led to significantly optimized circuit depths~\cite{schuster2025random}, enabling the practical implementation of random-state quantum algorithms on quantum hardware for tackling large-scale electronic-structure problems. The application of these quantum algorithms to realistic materials requires the availability of accurate real-space tight-binding (TB) Hamiltonians. Recently, classical machine-learning-based approaches have achieved high-fidelity TB parameterizations for complex materials~\cite{li2022deep,zhong2024universal,gu2024deep}, thereby bridging the gap between classical modeling and quantum computation and paving the way toward practical quantum simulations of large and complex material systems.

\begin{acknowledgments}
This work was supported by the Major Project for the Integration of Science, Education and Industry (Grant No. 2025ZDZX02). Feng Xiong acknowledges helpful discussions with Shan Jin. Xueheng Kuang thanks Xiaotian Yang, Yunhai Li and Weiqing Zhou for their fruitful discussions.
\end{acknowledgments}

\appendix

\section{Classical Method for Calculating the DOS Through TDPM}
\label{append_sec:DOS_te_app}

A practical classical method to compute the density of states (DOS) is the Chebyshev polynomial expansion of the time evolution operator. The Hamiltonian is first normalized as \(\widetilde{H} = \frac{H}{\alpha}\) so that\(\|\widetilde{H}\| \leq 1\). The propagator is then expanded as
\begin{equation}
    e^{-iHt} = J_0(\alpha t) + 2\sum_{m=1}^{\infty} (-i)^m J_m(\alpha t)\,T_m(\widetilde{H}),
\end{equation}
where \( J_m \) are Bessel functions and \( T_m \) are Chebyshev polynomials obeying the iterative relations
\begin{equation}
    T_0 = I,\quad T_1 = \widetilde{H},\quad T_m = 2\widetilde{H}T_{m-1} - T_{m-2}.
\end{equation}

To apply \( T_m(\widetilde{H}) \) on a state, one starts from a normalized random vector \( \ket{\psi^p_0} \) and generates iteratively
\begin{equation}
    \ket{r_m} = 2\widetilde{H}\ket{r_{m-1}} - \ket{r_{m-2}}.
\end{equation}
This avoids explicit matrix storage and yields
\begin{equation}
    e^{-iHt}\ket{\psi^p_0} = J_0(\alpha t)\ket{r_0} + 2\sum_{m=1}^\infty (-i)^m J_m(\alpha t)\ket{r_m}.
\label{eq:eiht_on_psi}
\end{equation}
The DOS is obtained from the Fourier transform of the averaged autocorrelation function
\begin{equation}
    C(t) = \lim_{S\to\infty}\frac{1}{S}\sum_{p=1}^S \braket{\psi_0^p|e^{-iHt}|\psi_0^p}.
\end{equation}
In practice, \( C(t) \) is sampled on a discrete grid and transformed using FFT\cite{yuan2010tbpm}. To reduce artifacts from finite cutoffs, a Hanning window function is applied before FFT, improving the smoothness of the DOS spectrum. 

\begin{figure}
    \centering
    \includegraphics[width=1.0\linewidth]{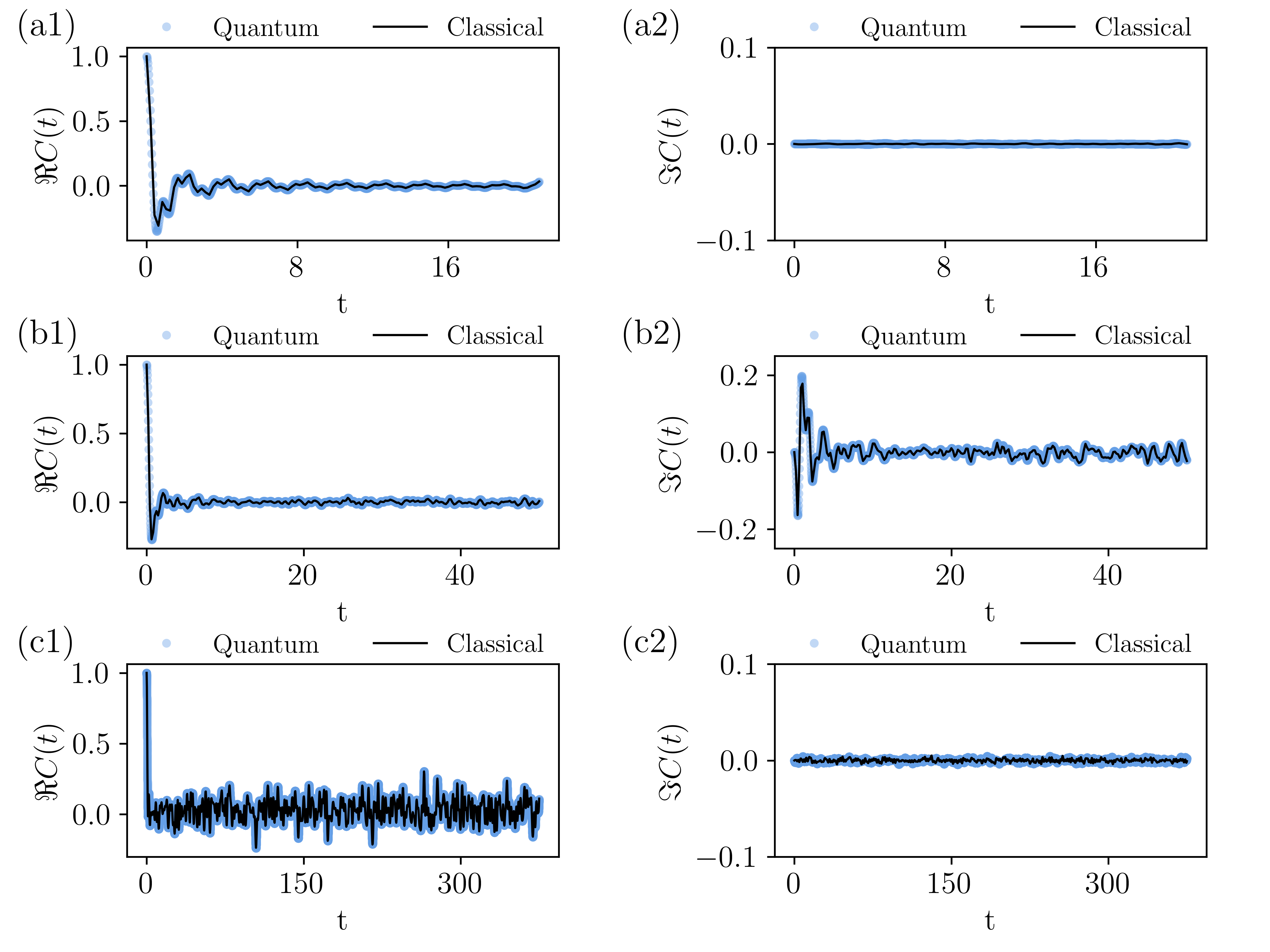}
    \caption{Comparison of time-dependent correlation functions $C(t)$ obtained from the classical TDPM (black solid lines) and the Q-TDPM simulations (blue circles). Panels (a1-c1) show the real parts of $C(t)$ and panels (a2-c2) show the imaginary parts for (a) the $64\times64$ graphene lattice, (b) the $30^\circ$-twisted bilayer graphene ($30^\circ$-TBG) disk (radius $4\,\text{nm}$, $3828$ atoms), and (c) the Sierpi\'{n}ski carpet fractal ($256$ atoms). }
    \label{fig:Corr}
\end{figure}
In our converged classical TDPM simulations, we matched the total evolution time and the number of random samples to those used in the Q-TDPM simulations for each system. Specifically, for the $64 \times 64$ graphene lattice and the 30$^\circ$-twisted bilayer graphene (30$^\circ$-TBG) disk (radius 4 nm, 3828 atoms), the classical TDPM adopted a time step 10 times larger than that of Q-TDPM and used one-tenth the number of steps. For the Sierpiński carpet fractal (256 atoms), the time step was increased by a factor of 32, with the number of steps reduced accordingly. These settings ensure that the total evolution time remains consistent with Q-TDPM. As a convergence check, Fig.~\ref{fig:Corr} shows that the correlation function computed by Q-TDPM using a much smaller Trotter step agrees well with the converged classical TDPM result.

\begin{figure}
    \centering
    \includegraphics[width=0.5\linewidth]{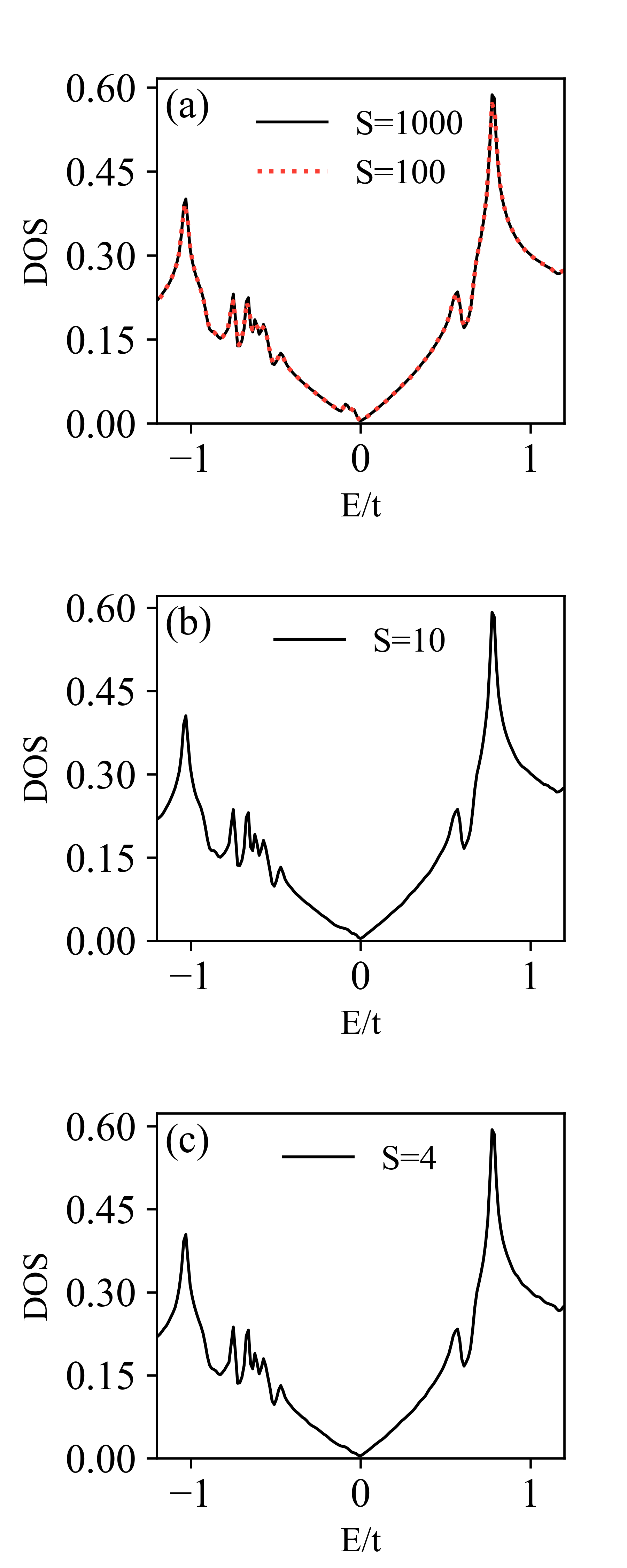}
    \caption{DOS of 30°-TBG with different disk radii calculated using the classical time-dependent propagation method (TDPM). (a) DOS for a disk of radius 20 nm computed with \(S = 100\) random samples (red dashed line) and \(S = 1000\) random samples (solid black line). (b) DOS for a disk of radius 46 nm computed with \(S = 10\) random samples. (c) DOS for a disk of radius 66 nm computed with \(S = 4\) random samples. }
    \label{fig:DOS_cry}
\end{figure}

We also clarify the finite-size effects that arise when simulating an ideal infinite quasicrystal, as discussed in the main text. For the infinite 30$^\circ$-twisted bilayer graphene (30$^\circ$-TBG), we perform classical TDPM calculations on circular flakes of varying radii. As shown in Fig.~\ref{fig:DOS_cry}(a), DOS computed on a relatively small 20 nm-radius disk (significantly larger than the Q-TDPM sample) already converges numerically with 100 random samples. Increasing the number of samples to 1000 does not restore the Dirac cone near zero energy that is known to exist in 30$^\circ$-TBG~\cite{ahn2018dodecagonal,yu2019dodecagonal}. However, this Dirac cone feature does re-emerge as the sample size increases from 46 nm to 66 nm, even with just a few random samples, as shown in Fig.~\ref{fig:DOS_cry}(b–c).These observations indicate that while increasing the number of random samples $S$ in the random-state method can reduce the stochastic trace estimation error to $\mathcal{O}(1/\sqrt{SN_{\text{shots}}})$~\cite{jin2021randomstate}, it cannot eliminate finite-size effects when approximating infinite systems. Proper recovery of the physical spectral features therefore requires sufficiently large system sizes, in addition to adequate sampling.

\section{Mapping the Time Evolution Operator $e^{-iHt}$ to a Quantum Circuit via Trotter Decomposition}
\label{append_sec:eiHt_to_cir}
\begin{figure}
    \centering
    \includegraphics[width=1\linewidth]{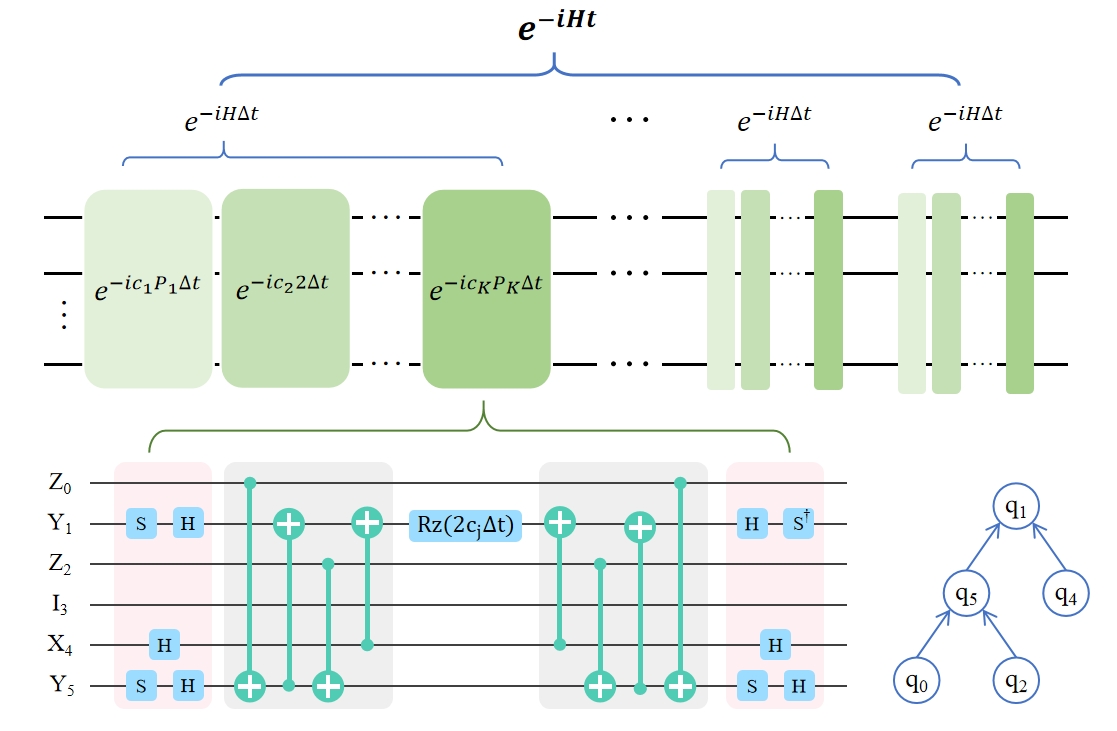}
    \caption{Example of implementing \(e^{-iHt}\) as a quantum circuit. The operator is approximated by M identical \(e^{-iH\Delta t}\) blocks via first-order Trotter-Suzuki decomposition. Each block is further decomposed into \(e^{-i c_j P_j \Delta t}\) operators, which are then mapped to quantum gates. Taking \(e^{-ic_j Y_5\otimes X_4 \otimes I_3 \otimes Z_2 \otimes X_1 \otimes Z_0 \Delta t}\) as an example, qubits with X or Y operators are initialized with H gates or S-H sequences (pink boxes), while Gray boxes represent CNOT gates entangling all non-identity qubits, forming a tree structure where all relevant qubits ultimately connect to the root qubit \(q_1\), where the rotation \(Rz(2c_j\Delta t)\) is applied. Inverse CNOT gates and final \(H/HS^\dagger\) gates on X/Y-qubits complete the interaction.}
    \label{fig:trotter}
\end{figure}
The material Hamiltonian is first decomposed into a summation of Pauli tensor products
\begin{equation}
    H = \sum_j c_j P_j,
\end{equation}
where $P_j = \bigotimes_{i=0}^{n-1} \sigma_i$ and $ \sigma_i \in \{I, X, Y, Z\} $. The time-evolution operator \( e^{-iHt} \) is then approximated via a first-order Trotter-Suzuki decomposition
\begin{equation}
    e^{-iHt} \approx \left( \prod_j e^{-ic_j P_j \Delta t} \right)^M,
\end{equation}
where \( \Delta t = t/M \) and \( M \) denotes the number of decomposition steps, defining the temporal resolution of the simulation. Each term \( e^{-ic_j P_j \Delta t} \) is then mapped to a quantum circuit. For basis transformations, apply Hadamard operators on qubits with operator $X$ and sequential \( S \) gate and \( H \) gate for \( Y \)-terms to align rotations with the \( Z \)-axis. Subsequently, all non-identity operator qubits are connected via CNOT gates, treating control qubits as arrow tails and target qubits as arrowheads. An unique core qubit is directly or indirectly targeted by all other non-identity qubits. A rotation gate \( R_z(2c_j\Delta t) \) is then applied to this core qubit. The previously applied CNOT gates are mirrored to restore the original topology. Finally, basis transformations are reverted. Apply Hadamard gates  for \( X \)-terms, and sequential \( H \) gate and  \( S^{\dagger} \) gate  for \( Y \)-terms. Fig.~\ref{fig:trotter} illustrates this protocol for the operator \( e^{-ic_j Y_5 X_4 \otimes I_3 \otimes Z_2 \otimes Y_1 \otimes Z_0 \Delta t} \).

\section{Classical Kernel Polynomial Method to approximate the DOS}
\label{append_sec:DOS_kpm_classical}
\begin{figure}
    \centering
    \includegraphics[width=1.0\linewidth]{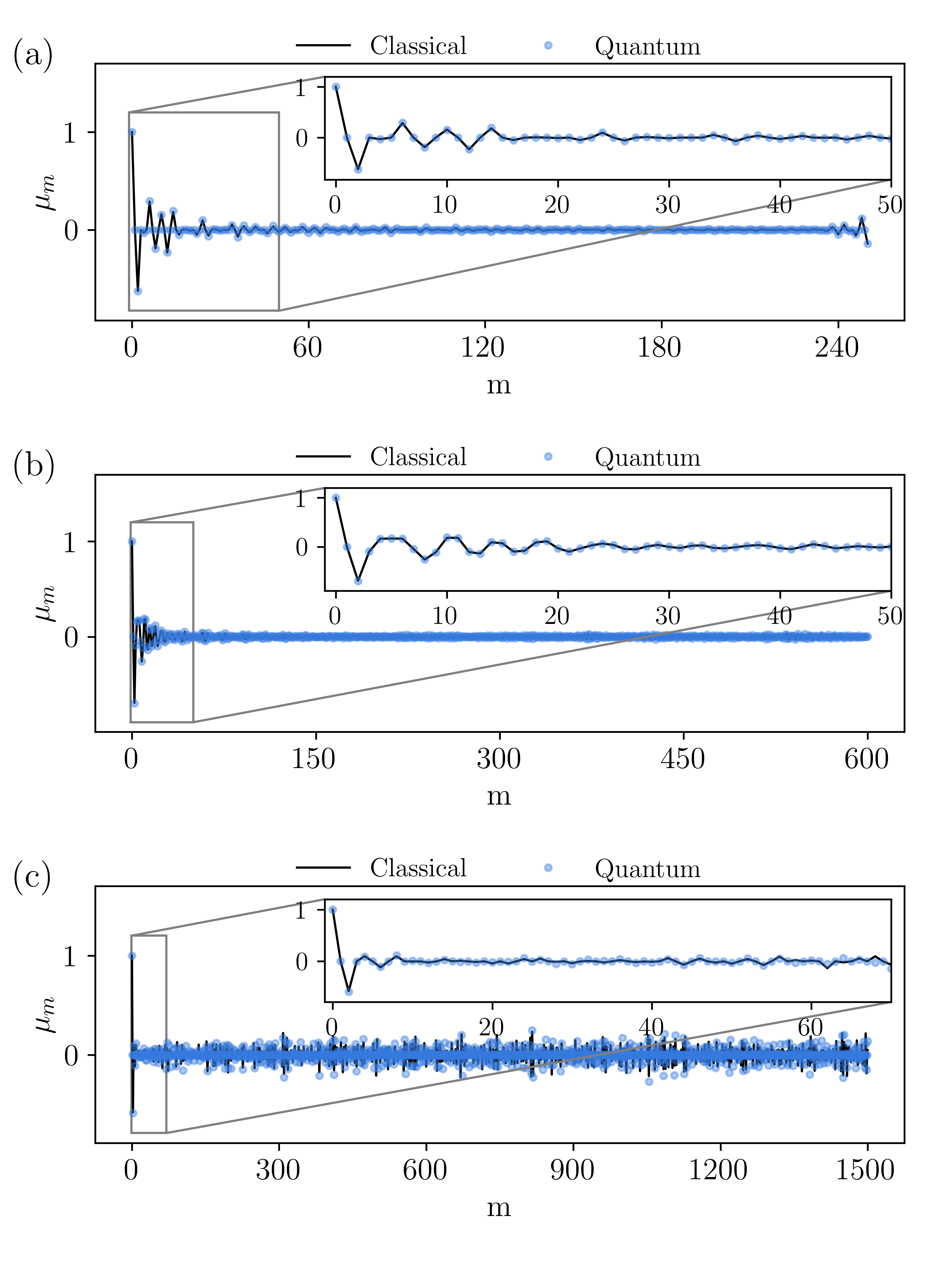}
\caption{Comparison of Q-KPM (blue circles) and classical KPM (black solid lines) results for the Chebyshev moments $\mu_{m}$ as a function of the moment index $m$, calculated using 1000 random samples. Panels (a), (b), and (c) correspond to a $64 \times 64$ graphene lattice, a 4-nm-radius disk of $30^{\circ}$-twisted bilayer graphene (30$^{\circ}$-TBG, 3828 atoms), and a Sierpiński carpet fractal (256 atoms) respectively.}
 \label{fig:mu}
\end{figure}

In contrast to the quantum formulation discussed in Sec.~\ref{subsec:DOS_kpm}, the classical KPM differs only in the evaluation of the Chebyshev moments \(\mu_m\). Specifically, they are defined as
\begin{equation}
    \mu_m = \frac{1}{N}\operatorname{Tr}\!\left[T_m(\widetilde{H})\right],
\end{equation}
which can be estimated stochastically by averaging over \(S\) random probe states:
\begin{equation}
    \mu_m \;\approx\; \frac{1}{S}\sum_{p=1}^{S}
    \braket{\psi^{p}_0 \mid T_m(\widetilde{H}) \mid \psi^{p}_0}.
\end{equation}

As described in the Eq.~(\ref{eq:eiht_on_psi}) of Appendix~\ref{append_sec:DOS_te_app} , the action of \(T_m(\widetilde{H})\) on a vector can be computed recursively without explicitly forming the polynomial, and the moments follow as
\begin{equation}
    \mu_m \;\approx\; \frac{1}{S}\sum_{p=1}^{S}
    \braket{\psi^{p}_0 \mid r^{p}_{m}}.
\end{equation}

Except for this difference in the computation of $\mu_m$, all other steps of the KPM procedure, including kernel damping, number of Chebyshev moments and the same rescaling factor for the Hamiltonian, are identical to those described in the quantum version in the main text.As shown in Fig.~\ref{fig:mu}, the Chebyshev moments obtained from Q-KPM (blue circles) closely follow those from classical KPM (black solid lines) for all three systems. This good agreement demonstrates that the Trotterized implementation of $e^{i m H_L}$ in Q-KPM can reproduce the converged classical KPM results to high precision.
\section{Classical Method for Calculating Quasi-eigenstates}
\label{append_sec:quasi_eig_app}

The quasi-eigenstate corresponding to an energy $\varepsilon$ can be formally defined as
\begin{equation}
\ket{\Psi(\varepsilon)} = \frac{1}{2\pi} \int_{-\infty}^{\infty} e^{i\varepsilon t} \ket{\psi(t)} \, dt,
\end{equation}
where the time-dependent state $\ket{\psi(t)} = e^{-iHt}\ket{\psi_{0}}$ follows the same Chebyshev expansion and recursive procedure as described in Eq.~(\ref{eq:eiht_on_psi}) of Appendix~\ref{append_sec:DOS_te_app}. In numerical practice, the continuous integral is replaced by a discrete Fourier transform, efficiently implemented using FFT, so that $\ket{\Psi(\varepsilon)}$ can be obtained from a time series of $\ket{\psi(t)}$ evaluated at evenly spaced time steps.

In $\ket{\Psi(\varepsilon)}$, each component corresponds to a specific lattice site in the Hamiltonian basis. Because the Hamiltonian is constructed by ordering the basis states according to the actual spatial arrangement of the system, the $i$-th element $\Psi_{i}(\varepsilon)$ directly maps onto the $i$-th real-space site. Taking the squared modulus $|\Psi_{i}(\varepsilon)|^{2}$ then yields the local probability density of the quasi-eigenstate at that site. In this way, the entire state vector $\ket{\Psi(\varepsilon)}$ naturally represents the spatial distribution of the system. For visualization, the known spatial coordinates of each lattice site are used to project the probability densities $|\Psi_{i}(\varepsilon)|^{2}$ onto the corresponding real-space positions. The resulting values can be represented by color scales or marker sizes to produce a spatial map of the quasi-eigenstate across the system.

In M-QPE, the measured probabilities of the computational basis states correspond one-to-one to $|\Psi_{i}(\varepsilon)|^{2}$, and the indexing of the basis states follows the same order as the lattice sites used in the Hamiltonian construction. Consequently, the same coordinate mapping can be applied to obtain the real-space distribution, making the classical and quantum results directly comparable.
In our calculations, the time step $\Delta t$ and the number of steps are chosen to be the same as those used in the quantum algorithm described in Sec.~\ref{subsec:quasieigen}, ensuring a consistent frequency resolution and directly comparable results between the classical and quantum approaches. The M-QPE algorithm implements the time evolution at each step via a Trotter decomposition of $e^{-iHt}$, whereas the classical calculation uses the Chebyshev recursion; both methods converge to the same quasi-eigenstate.

\section{Atomic structrue}
\label{append_sec: Atomicstructure}

\begin{figure}
    \centering
    \includegraphics[width=1\linewidth]{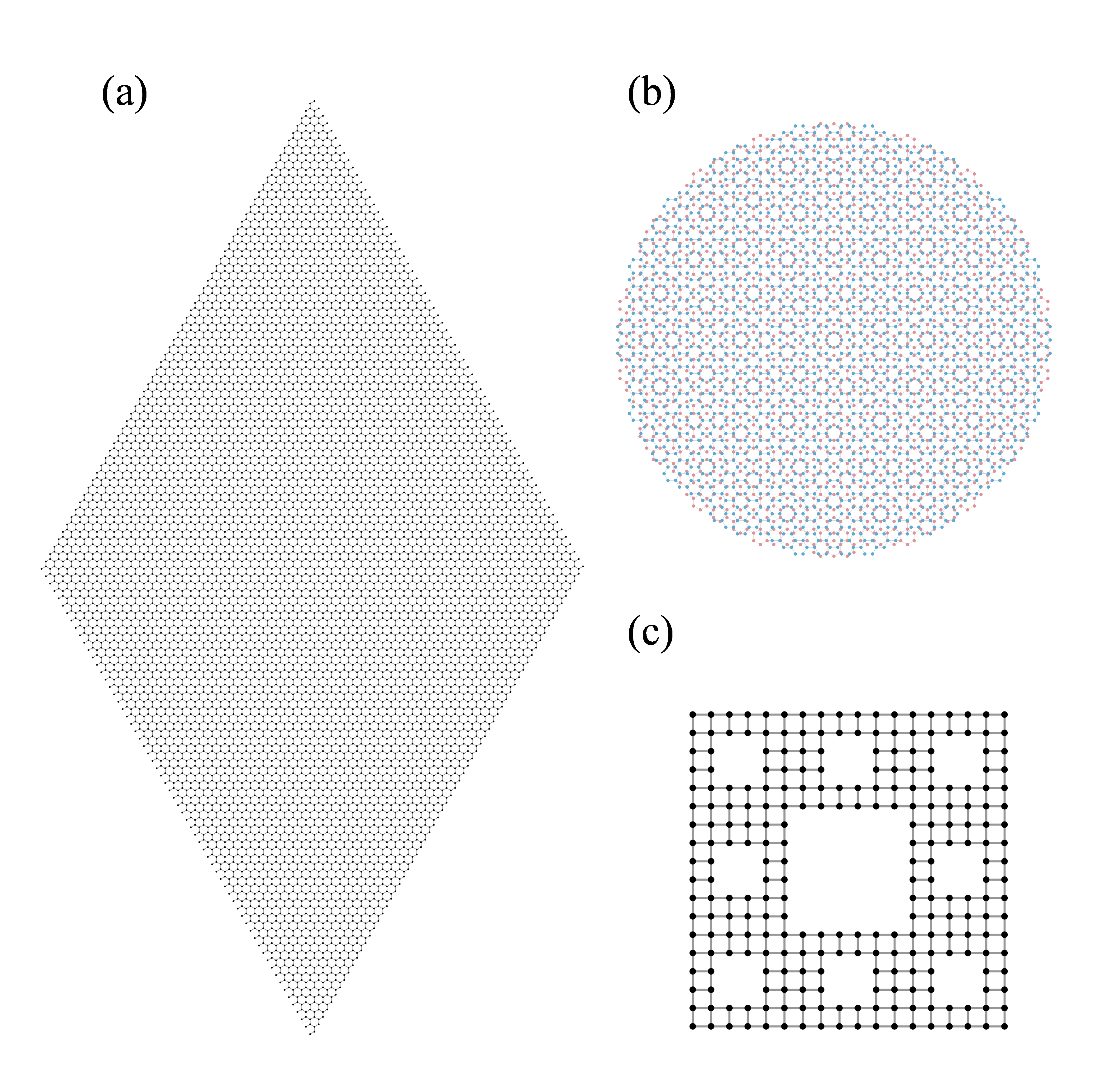}
    \caption{(a) Graphene supercell with 4096 carbon atoms, (b) 30$^\circ$-TBG quasicrystal, and (c) fractal samples for DOS calculations. The carbon atoms are represented by lattice points.}
    \label{fig:atomicstructure}
\end{figure}

The atomic structure of periodic graphene shown in Fig.~\ref{fig:atomicstructure}(a) was constructed from a hexagonal graphene primitive cell with a lattice constant \( a = 2.46~\text{\AA} \). The 30$^\circ$-twisted bilayer graphene (30$^\circ$-TBG) quasicrystal shown in 
Fig.~\ref{fig:atomicstructure}(b) was generated by rotating two layers of graphene with an interlayer distance \( d = 3.35~\text{\AA} \). 
A top-down approach was employed to build the Sierpiński carpet fractal shown in Fig.~\ref{fig:atomicstructure}(c). 
All atomic structure samples are accessible through TBPLaS~\cite{li2023tbplas}.

\section{Real-Space Atomic Tight-Binding Hamiltonian}
\label{append_sec:TBmodel}

In the main text, we adopt an atomic tight-binding (TB) Hamiltonian to describe the electronic structure of both periodic and non-periodic graphene-based systems. The model focuses on the $p_z$ orbital of each carbon atom, which governs the low-energy physics of graphene. The Hamiltonian is expressed as
\begin{equation}
H = \sum_i \epsilon_i c_i^{\dagger} c_i + \sum_{i \neq j} t_{ij} c_i^{\dagger} c_j,
\label{hal0}
\end{equation}
where $\epsilon_i$ denotes the on-site energy of the $p_z$ orbital at site $i$, and $t_{ij}$ represents the hopping amplitude between orbitals at sites $i$ and $j$.

For monolayer graphene, only the nearest-neighbor hopping is considered, with $t_{ij} = t = -2.7$~eV~\cite{neto2009graphene} and $\epsilon_i = 0$. In the case of the Sierpiński fractal lattice, we use $t = -1$~eV~\cite{yang2020confined}. For the 30$^\circ$-TBG quasicrystal, the hopping amplitudes are computed using the Slater--Koster (SK) formalism:
\begin{equation}
t_{ij} = n^2 V_{pp\sigma}(r_{ij}) + (1 - n^2) V_{pp\pi}(r_{ij}),
\label{eq:tij}
\end{equation}
where $r_{ij} = |\mathbf{r}_j - \mathbf{r}_i|$ is the distance between atoms $i$ and $j$, and $n$ is the direction cosine along the $\mathbf{e}_z$ axis, normal to the graphene plane. The SK parameters are given by references~\cite{trambly2010localization,trambly2012numerical}:
\begin{align}
V_{pp\pi}(r_{ij}) &= -t_0 \exp\left[q_{\pi} \left(1 - \frac{r_{ij}}{d}\right)\right] F_c(r_{ij}), \label{eq:vpppi} \\
V_{pp\sigma}(r_{ij}) &= t_1 \exp\left[q_{\sigma} \left(1 - \frac{r_{ij}}{h}\right)\right] F_c(r_{ij}), \label{eq:vppsigma}
\end{align}
where $d$ and $h$ denote the in-plane and out-of-plane nearest-neighbor carbon--carbon distances, respectively. We use $t_0 = 2.7$~eV and $t_1 = 0.48$~eV to set the respective hopping strengths. The decay factors satisfy the condition
\[
\frac{q_{\sigma}}{h} = \frac{q_{\pi}}{d} = 2.218~\text{\AA}^{-1}.
\]

To smoothly suppress long-range interactions, we apply the cutoff function
\begin{equation}
F_c(r) = \frac{1}{1 + \exp\left[(r - r_c)/l_c\right]},
\end{equation}
with decay length $l_c = 0.265$~\AA{} and cutoff distance $r_c = 7.5$~\AA.

\bibliography{reference.bib} 
\end{document}